\documentstyle[11pt,aaspp4]{article}

\slugcomment{ApJ (Main Journal); accepted September 3, 1997}

\begin{document}

\oddsidemargin  -0.5pc
\evensidemargin -0.5pc

\def\abs#1{\left| #1 \right|}
\def\EE#1{\times 10^{#1}}
\def\gcm{\rm ~g~cm^{-3}}
\def\cm3{\rm ~cm^{-3}}
\def\kms{\rm ~km~s^{-1}}
\def\cms{\rm ~cm~s^{-1}}
\def\ergs{\rm ~ergs~s^{-1}}
\def\isotope#1#2{\hbox{${}^{#1}\rm#2$}}
\def\wl{~\lambda}
\def\wll{~\lambda\lambda}
\def\HI{{\rm H\,I}}
\def\HII{{\rm H\,II}}
\def\HeI{{\rm He\,I}}
\def\HeII{{\rm He\,II}}
\def\HeIII{{\rm He\,III}}
\def\CI{{\rm C\,I}}
\def\CII{{\rm C\,II}}
\def\CIII{{\rm C\,III}}
\def\CIV{{\rm C\,IV}}
\def\NI{{\rm N\,I}}
\def\NII{{\rm N\,II}}
\def\NIII{{\rm N\,III}}
\def\NIV{{\rm N\,IV}}
\def\NV{{\rm N\,V}}
\def\NVI{{\rm N\,VI}}
\def\NVII{{\rm N\,VII}}
\def\OI{{\rm O\,I}}
\def\OII{{\rm O\,II}}
\def\OIII{{\rm O\,III}}
\def\OIV{{\rm O\,IV}}
\def\OV{{\rm O\,V}}
\def\OVI{{\rm O\,VI}}
\def\OVII{{\rm O\,VII}}
\def\OVIII{{\rm O\,VIII}}
\def\CaI{{\rm Ca\,I}}
\def\CaII{{\rm Ca\,II}}
\def\NeI{{\rm Ne\,I}}
\def\NeII{{\rm Ne\,II}}
\def\NeIII{{\rm Ne\,III}}
\def\NeIV{{\rm Ne\,IV}}
\def\NeV{{\rm Ne\,V}}
\def\NaI{{\rm Na\,I}}
\def\NaII{{\rm Na\,II}}
\def\NiI{{\rm Ni\,I}}
\def\NiII{{\rm Ni\,II}}
\def\FeI{{\rm Fe\,I}}
\def\FeII{{\rm Fe\,II}}
\def\FeIII{{\rm Fe\,III}}
\def\FeV{{\rm Fe\,V}}
\def\FeVII{{\rm Fe\,VII}}
\def\CoII{{\rm Co\,II}}
\def\CoIII{{\rm Co\,III}}
\def\ArI{{\rm Ar\,I}}
\def\MgI{{\rm Mg\,I}}
\def\MgII{{\rm Mg\,II}}
\def\SiI{{\rm Si\,I}}
\def\SiII{{\rm Si\,II}}
\def\SiIII{{\rm Si\,III}}
\def\SiIV{{\rm Si\,IV}}
\def\SiVI{{\rm Si\,VI}}
\def\SI{{\rm S\,I}}
\def\SII{{\rm S\,II}}
\def\SIII{{\rm S\,III}}
\def\SIV{{\rm S\,IV}}
\def\SVI{{\rm S\,VI}}
\def\FeI{{\rm Fe\,I}}
\def\FeII{{\rm Fe\,II}}
\def\FeIII{{\rm Fe\,III}}
\def\FeIV{{\rm Fe\,IV}}
\def\FeVII{{\rm Fe\,VII}}
\def\kI{{\rm k\,I}}
\def\kII{{\rm k\,II}}
\def\La{{\rm Ly}\alpha}
\def\Ha{{\rm H}\alpha}
\def\Hb{{\rm H}\beta}
\def\Lya{{\rm Ly}\alpha}
\def\etscale#1{e^{-t/#1^{\rm d}}}
\def\etscaleyr#1{e^{-t/#1\,{\rm yr}}}
\def\sigmaKN{\sigma_{\rm KN}}
\def\ncrit{n_{\rm crit}}
\def\Emax{E_{\rm max}}
\def\chieff{\chi_{\rm eff}^{\phantom{0}}}
\def\chieffi{\chi_{{\rm eff},i}^{\phantom{0}}}
\def\chiion{\chi_{\rm ion}^{\phantom{0}}}
\def\chiioni{\chi_{{\rm ion},i}^{\phantom{0}}}
\def\Gammaion{\Gamma_{\!\rm ion}}
\def\Mcore{M_{\rm core}}
\def\Rcore{R_{\rm core}}
\def\Vcore{V_{\rm core}}
\def\Menv{M_{\rm env}}
\def\Venv{V_{\rm env}}
\def\Vej{V_{\rm ej}}
\def\Vcthou{\left( {\Vcore \over 2000 \rm \,km\,s^{-1}} \right)}
\def\KK{\rm ~K}
\def\Msun{~M_\odot}
\def\Rsun{~R_\odot}
\def\Msunyr{~M_\odot~{\rm yr}^{-1}}
\def\Mdot{\dot M}
\def\M56{M_{\rm ej,56}}
\def\MZA{M_{\rm ZAMS}}
\def\tyr{t_{\rm yr}}
\def\gff{g_{\rm ff}}
\def\Tex{T_{\rm ex}}
\def\math#1{\vskip20true pt\hskip3true cm{#1}}
\def\no{\hang\noindent}
\def\dots{$\ldots$}
\def\etal{{\it et al.}}
\def\ie{{\it i.~e.\ }}
\def\ien{{\it i.~e.~}}
\def\p{\partial}
\def\pp#1{{\partial \over {\partial{#1}}}}
\def\der#1{{d \over {d{#1}}}}
\def\bigskip{\vskip1true cm}
\def\eg{e.~g.~}
\def\prop{\propto}

\title{A very low mass of $^{56}$Ni in the ejecta of SN 1994W}

\author{Jesper Sollerman\altaffilmark{1}, Robert J. Cumming\altaffilmark{1} and 
Peter Lundqvist\altaffilmark{1,2}}

\altaffiltext{1}{Stockholm Observatory, SE-133 36 Saltsj\"obaden, Sweden.} 
\altaffiltext{2}{Send offprint requests to Peter Lundqvist; E-mail: peter@astro.su.se}

\begin{abstract}
We present spectroscopic and photometric observations of the luminous
narrow-line Type IIP (plateau) supernova 1994W.  After the plateau
phase ($t \gtrsim 120$~days), the light curve dropped by $\sim 3.5$~mag 
in $V$ in only 12 days.  Between 125 and 197 days
after explosion the supernova faded substantially faster than the
decay rate of $^{56}$Co, and by day 197 it was 3.6~magnitudes less luminous 
in $R$ compared to SN 1987A. The low $R$-luminosity could 
indicate $\lesssim 0.0026^{+0.0017}_{-0.0011} \Msun$ of $^{56}$Ni ejected at 
the explosion, but the emission between 125 and 197 days must then have been
dominated by an additional power source, presumably circumstellar interaction.
Alternatively, the late light curve was dominated by $^{56}$Co decay. In this
case, the mass of the ejected $^{56}$Ni was $0.015^{+0.012}_{-0.008}\ \Msun$,
and the rapid fading between 125 and 197 days was most likely due to dust 
formation. Though this value of the mass is higher than in the case with the
additional power source, it is still lower than estimated for any previous 
Type II supernova. 

Only progenitors with $\MZA \sim 8-10 \Msun$ and $\MZA \gtrsim 25 \Msun$ 
are expected to eject such low masses of $^{56}$Ni. 
If $\MZA \sim 8-10 \Msun$, the plateau phase indicates a low explosion energy,
while for a progenitor with $\MZA \gtrsim 25 \Msun$ the energy can be the 
canonical $\sim 10^{51}$~ergs. As SN 1994W was unusually luminous, the low-mass
explosion may require an uncomfortably high efficiency in converting explosion
energy into radiation. This favors a $\MZA \gtrsim 25 \Msun$ progenitor. 

The supernova's narrow ($\sim 1000 \kms$) emission lines were excited by the 
hot supernova spectrum, rather than a circumstellar shock. The thin shell from
which the lines originated was most likely accelerated by the radiation from 
the supernova.
\end{abstract}


\section{Introduction}
Type II supernovae (SNe II) are thought to arise from the explosion of
massive ($\MZA \gtrsim 8-10 \Msun$) stars which end their lives with a
substantial fraction of hydrogen in their envelopes. The explosion is 
triggered by the collapse of the progenitor's core. The best-understood
core-collapse events are the SNe IIP (plateau), whose light curves are 
nearly flat during the first $\sim 2 - 4$ months. The long duration of 
the light curve plateau is believed to be due to the presence of a massive
($\gtrsim 2-3 \Msun$) hydrogen envelope (e.g., Falk $\&$ Arnett 1977).

After the plateau, the light curves of SNe IIP are often dominated by 
radioactive decay of $^{56}$Co. The cobalt is the decay product of $^{56}$Ni,
and the luminosity at this epoch can therefore be used to estimate the amount 
of $^{56}$Ni ejected at the explosion. Because the nickel is produced in the 
core region, the mass of ejected $^{56}$Ni is an important probe of the 
core-collapse scenario itself, the nucleosynthetic yields of SNe II and their
progenitors, and the nature of the compact central object formed at the 
collapse (\cite{WW95}; \cite{WT96}).

A new subclass of core-collapse supernovae has recently emerged: the 
narrow-line supernovae or SNe IIn (Schlegel 1990; Filippenko 1997). Like 
other SNe II they show strong Balmer lines, but the lines are narrow
($\sim 1000-3000 \kms$) and are seen in emission rather than absorption even 
at early times.

SN 1994W was discovered on 1994 July 29 in the Sbc galaxy NGC 4041 
(\cite{CV94}). Based on the presence of hydrogen lines it was quickly
classified as a SN II (Bragaglia, Munari, \& Barbon 1994).
Subsequent spectroscopy showed that the lines had unusually
narrow P-Cygni profiles ($v_{\rm FWHM} \sim 1200 \kms$) with broad
emission wings, but no broad absorption component (\cite{FB94};
Cumming, Lundqvist, \& Meikle 1994).  Reports in IAU circulars
indicated that the supernova reached a maximum of $V \sim 13.3$ around
1994 Aug 13.  Further photometry by Tsvetkov (1995) showed that the supernova
was a SN IIP, and had a remarkably sharp drop in the $B$ and $V$ light curves 
in early November 1994.

In this paper we present further photometric and
spectroscopic observations of SN 1994W taken on several occasions
between 1994 July 31 to 1996 April 20. We focus on the low luminosity of the
supernova after $\sim 120$ days, and interpret this as due to an exceptionally
low mass of ejected $^{56}$Ni. We also discuss the formation of the narrow
lines.

\section{Observations}
\subsection{Photometry}
Our photometric data were obtained mainly in $R$ on six occasions from 
1994 November 16 to 1996 April 20, using the Jakobus Kapteyn Telescope
(JKT) and the
Nordic Optical Telescope (NOT) on La Palma (see Table~1). The data were bias 
subtracted and flat fielded
using standard procedures within IRAF.
Differential magnitudes between
the supernova and comparison stars were
measured using POLYFIT (Sp\"annare 1997), which is a software package 
developed to do photometry on uneven backgrounds: 
a point spread function (PSF) is produced by fitting a Moffat function 
to an isolated star in the frame. This PSF and a polynomial fit to the 
background are then simultaneously fitted at the position of the
object. For the JKT nights a bright comparison star was in the
field and  was
used for the relative magnitudes. For the NOT data a fainter
comparison star had to be used.  Up to seven comparison stars were
monitored to avoid
variability.
The comparison stars were calibrated against Landolt stars 
(Landolt 1983, 1992) at two nights on JKT and against a standard field
in NGC 4147 
(Christian et al. 1985) for the NOT observations. The zero point for the $R$
magnitude differs slightly between the 4 photometric
nights. We estimate a total systematic error of 0.15 mag in the $R$
magnitude, including the neglect of color transformations and differential 
atmospheric extinction
corrections, which were only applied  to the day 646\footnotemark\  
\footnotetext{Days are measured after 14.0 July, 1994 (see \S 3)} data.
For the $V$ magnitude we calibrated relative to local standards
measured by Tsvetkov (1995).

The JKT observations have rather low signal to noise or poor seeing, 
which gives large errors for the photometry, $\sim \pm$0.5 mag. For the NOT 
observation (day 197) the error is smaller, and dominated by the error in 
the POLYFIT measurement. We estimated this error by adding to the image 
ten artificial stars of the same magnitude as the supernova at similar 
backgrounds using the IRAF task ADDSTAR. The rms error in the POLYFIT 
measurements of these were $\lesssim 0.15$~mag, which we regard as the 
measurement error also for the supernova. A total error of 0.3 mag on day 197 
is therefore rather generous and includes errors for the comparison stars. 
At the last observation, day 646, the supernova was not detected. We estimated 
an upper limit by subtracting artificial PSFs of different brightness
until a hole appeared in the background at the supernova position.

\subsection{Spectroscopy}
Our spectroscopic observations cover five epochs from day 18 to day 202 
(see Table~2). The data were bias subtracted, flat fielded and wavelength
calibrated using IRAF. The day 31, 57 and 202 spectra were flux calibrated 
against a spectrophotometric standard star. 
The day 57 spectrum was taken at high airmass and required further 
corrections for slit losses. A rough flux calibration for the 
day 18 spectrum was obtained by comparison with contemporary visual photometry
and the spectrum on day 31.  The day 121 spectrum was flux calibrated using 
the $R$ photometry from day 125.

\section{Results}
In Figure 1 we present the light curve of SN 1994W. 
From the early $R$ data, we estimate the date of optical outburst to have 
been 1994 July 14$^{+2}_{-4}$, and we label observations 
by number of days after this.  The light curve shows
a relatively gentle rise to maximum around 30 days, followed by a slow, 
plateau-like decline until $\sim 110$ days. The slow decline in $B$
and $V$ is typical of a SN IIP.
At $\sim 110$ days, however, the supernova started to fade unprecedentedly 
fast, dropping by $\sim 3.5$ magnitudes in $V$ in only 12 days. Also on
the light curve tail, the fading continued at a fast rate, until 
our last photometric detection on day 197. We could not detect the
supernova in $V$ or $R$ on day 646.

The spectrum of the supernova was also unusual for a SN II.
Before the light curve drop, the spectra on days 18, 31 
and 57 (see Figs. 2 and 3), show the strong P-Cygni
lines of $\HI$, typical of a SN II, but with very narrow profiles 
(minima at $-700 \kms$ and maximum blue velocities of only $1000 \kms$).
In addition to the narrow features, the $\HI$ lines had
emission wings extending out to $\sim 5000 \kms$, a factor of $\sim 2-4$ 
lower than the maximum velocities normally seen in SNe II at these epochs.  
The $\HI$ lines were accompanied by a rich spectrum of narrow lines of 
low-ionization species (He~I, O~I, Mg~II, Si~II, Fe~II).  
Though conspicuous, the emission lines account
for only $\sim 1\%$ of the total optical flux on days 31 and 57.
Instead, a blue continuum dominates. Estimating $E(B-V) = 0.17\pm0.06$ from
the interstellar Na~I~D absorption in our day 31 spectrum (\cite{MZ97}), we fit 
it with blackbody spectra with temperatures of $\sim (1.50\pm0.15)\EE4$~K 
and $\sim (1.10\pm0.10)\EE4$~K, respectively.
The effective temperatures indicated by the blackbody fits are much higher
than the $\sim (5-8) \EE3$~K usually observed in SNe IIP at similar epochs. 

The decrease in
temperature was accompanied by the disappearance of the $\HeI$ lines
and increasingly strong $\FeII$ features.  A spectrum taken by
Filippenko (1997) on day 80 shows that the spectral features had
changed only marginally compared to our day 57 spectrum, but the slope
of the continuum indicated a decrease in
effective temperature to $\sim (7-8)\EE3$~K.

The photometric fading at $\sim$110 days coincided with dramatic changes in
the spectrum.  The $B-V$ color, previously steady at $0.3\pm0.1$, 
increased to 1.75 between days 85 and 111 (Fig.~1; \cite{T95}). In addition, 
the flux in $\Ha$ dropped at roughly the same rate as the $V$ and $R$ light
curves. This coincidence, together with
the shift to lower-excitation lines during the plateau phase, suggests
that the narrow lines were predominantly excited by the supernova
continuum rather than by a circumstellar shock (see \S 4.4).

The day 121 spectrum shows narrow ($\sim 730\pm120 \kms$) emission 
in $\Ha$, $\NaI$~D $\lambda$5893 and [Ca~II] $\lambda\lambda$7291, 7324,
on top of a continuum-like spectrum. The flux of the narrow $\Ha$ component 
had dropped by a factor $\sim 150$ since day 57. No broad features can be 
identified, but the sharp edge present at $\sim5600$ \AA\ is a common 
feature in late-time supernova spectra,
usually attributed to the redmost extent of blended
Fe~II or [Fe~II] lines.  We suggest that the apparent continuum is in
fact made up of overlapping broad emission lines, with a gap around
5650~\AA.  Unlike other Type II events at this epoch, there is no sign
of strong, broad $\Ha$.

On day 202, the spectrum was dominated by a fairly flat, noisy
continuum and unresolved $\Ha$, which may contain a
contribution from a background $\HII$ region. The flux in the narrow
$\Ha$ line had dropped by a factor of $\gtrsim 5$ compared to the day
121 spectrum, but its width remained constant within the errors. The
flux in the $R$ band of our spectrum on day 202 agrees well with our
$R$ magnitude on day 197.  Our photometric limit from day 646 ($R >
22$) demonstrates that the supernova continued to fade after day 202,
and gives a lower limit on the flux contribution of the supernova to
the day 202 spectrum of $\sim 50\%$.
We suggest that the continuum at this epoch comes from background stars.
We tentatively identify a 
broad, $v_{\rm FWHM}=(4.5\pm1.5)\times10^3 \kms$, 
emission feature at $\sim7350$~\AA\ as [Ca~II]~$\lambda\lambda$7291, 7324, 
and suggest that the remaining emission left after the continuum is 
subtracted is, as on day 121, composed of broad emission lines. The [Ca~II]
feature has luminosity a factor of $\sim$100 lower than for SN 1987A
at the same epoch; the luminosity of $\Ha$ is at least a factor of 2
lower still (Fig. 4).

\section{Discussion}
In Figure~5 we compare the absolute light curve of SN 1994W to those of other 
SNe IIP, SN 1987A and the luminous SN IIL (linear) 1979C. 
At maximum, SN 1994W was as luminous as SN 1983K, the most luminous
SN IIP previously observed (Phillips et al. 1990). Evident from Figure~5
is the rapid fading and low luminosity of SN 1994W after the plateau phase ends,
i.e., $\sim 110$~days, and the fast decline after that. We now investigate 
whether the unusual light curve and spectra of SN 1994W can be explained 
in terms of standard supernova theory.

\subsection{The mass of ejected $^{56}$Ni}
For a SN IIP, the mass of ejected $^{56}$Ni, $\M56$, can be estimated
from the luminosity after the plateau phase ends. At this point the optical
emission is powered by $\gamma-$ray energy deposited into the ejecta from 
the radioactive decay of $^{56}$Co, the decay product of $^{56}$Ni. 
The e-folding time for the cobalt decay is 111.26 days, 
so the radioactive decay creates a tail to the bolometric light curve with a 
slope of  0.00976 mag day$^{-1}$, if all the $\gamma-$rays are trapped. 
The luminosity of the tail gives a measure of $\M56$, 
when other contributions to the light curve have been accounted for.

The mass of ejected nickel has been determined in this way for a
handful of SNe IIP, for example SNe 1969L ($0.07\pm0.03 \Msun$, \cite{S93}),
1990E ($0.073^{+0.018}_{-0.051} \Msun$, \cite{S93}),
1992H ($\sim0.075 \Msun$, \cite{C96}) and 
SN 1991G ($0.024^{+0.018}_{-0.010} \Msun$, \cite{B95}).

Our method to estimate $\M56$ of SN 1994W is to compare the level of its 
luminosity in $R$ after the plateau to that of SN 1987A at similar epochs.
We choose SN 1987A as a template since its distance and 
extinction are well known, and because its light curve was entirely powered 
by radioactive decay as soon as a few days after explosion. 
SN 1987A also resembled a SN IIP in that it had a massive hydrogen 
envelope (\cite{W88}). For SN 1987A the value
of $\M56$ was $\sim0.071^{+0.019}_{-0.016} \Msun$ (\cite{SB90}).

A problem with our method is that the light curve tail of SN 1994W declines 
substantially faster than the decay rate of $^{56}$Co. We must therefore be
careful in spelling out the assumptions involved in interpreting the light
curve.  We identify two possible scenarios which can
lead to such a decline, and which lead to different estimates of
$\M56$.  Radioactive decay of $^{56}$Co could dominate, but be unable for
some reason to maintain the standard tail of the $R$ light curve.
Alternatively, some source other than $^{56}$Co decay dominates the light
curve during this period.  In the following sections we investigate
both these possibilities and their implications for $\M56$.


\subsection{The steep decline of the late light curve: implications for 
the nickel mass}
%
First we consider the possibility that all of the flux on the tail
was indeed due to cobalt decay.  The steep decline could then be due to 
increasing escape of $\gamma$-rays, a gradual shift of the spectrum 
from the optical to the infrared, or increasing dust obscuration. 

Assuming a central radioactive source, the optical depth to $\gamma$-rays 
through the ejecta decreases with time as $\tau_{\gamma} = (t_1/t)^2$, 
where $t_1$ is the time when $\tau_{\gamma} = 1$. 
In SN 1987A, $t_1 \sim 750$~days (Kozma 1996). 
Including the fact that $\sim 3.5$\% of the cobalt decay energy is mediated 
by positrons and therefore deposited locally, the bolometric luminosity
decays by the factor $f_{\rm leak} = (1 - 0.965~{\rm exp}(-\tau_{\gamma}))$ 
in addition to the decay of $^{56}$Co. The observed flux in $R$ between days 
125 and 197 drops by a factor $f_{\rm obs} = 6.6^{+7.2}_{-3.5}$ in addition 
to cobalt decay. With the assumptions above the maximum value of $f_{\rm leak}$
for the same period is only $2.0$ (for $t_1 \sim 77$~days) and thus cannot 
reproduce the fast decline of the light curve. 
We therefore believe that $\gamma$-ray leakage was not
the cause of the exceptionally fast fading of SN 1994W after $\sim 120$~days.

A gradual shift of the spectrum into the infrared might be expected if
the mass of $^{56}$Ni was substantially lower than in SN 1987A, 
and the reduced radioactive heating lowered the ejecta temperature. In this
scenario our assumption of linear scaling between $\M56$ and the $R$-luminosity 
could break down. To test this numerically, we ran two models of SN 1987A kindly
provided for us by C. Kozma. These models do not include dust formation and
are similar to her mixed 10H model in Kozma (1996), 
but with only $\M56 = 0.001$ $(0.01) \Msun$, and with other radioactive 
isotopes decreased accordingly. We find that on day 150, more than 60 (70)\% 
of the line emission comes out between $3000 - 9000$~\AA, $\sim 7.3 (5.3)\%$ 
and $\sim 6.6 (2.2)\%$ of which in H$\alpha$ and [Ca~II], respectively. 
Therefore, if dust formation is unimportant, and the structure similar to
that of SN 1987A, linear scaling of $\M56$ with $R$ and $V$ luminosities 
seems to be valid down to very low values of $\M56$. 

The situation is different if dust formation is important.
We expect that below some value of $\M56$, radioactive heating of the ejecta
will be unable to prevent the formation of dust. This could attenuate
and shift the supernova emission into the infrared already shortly after
the plateau phase. The effect could be further amplified by increased 
molecular cooling, since the destruction rate of molecules will decrease 
with lower $\M56$. We know that dust increased the cooling and obscured the
line-emitting region in SN 1987A (Kozma 1996, and references therein), 
albeit at a later epoch (between $\sim 350 - 600$~days) than
considered here for SN 1994W. The dust covering factor would have increased
with time, creating a blueshift of the supernova lines, as observed for SN
1987A (Danziger et al. 1991). Our day 202 spectrum is too noisy to reveal
such an effect. Thus, we cannot rule out dust formation as the cause of the
fast fading in $R$.

To obtain an estimate of $\M56$ in the case of early dust formation, 
we choose the $R$ measurement from day 139, which is the first photometric 
point on the light curve tail. 
This gives $\M56 = 0.015^{+0.012}_{-0.008}\ \Msun$. The estimated error 
includes the statistical error shown in Figure~1, uncertainty in the 
distance modulus of SN 1994W ($\pm$0.3 mag), the extinction towards the 
supernova ($\pm$0.15 mag) and an estimated systematic error of our $R$ 
calibration ($\pm$0.15 mag). 

This $\M56$ is slightly lower than for the previous record-holder, SN
1991G, which had $0.024^{+0.018}_{-0.010} \Msun$. However, SN 1991G did
not show the very steep decline of the light curve tail (\cite{B95}). 
This could mean that even SN 1991G ejected too much $^{56}$Ni to allow early
dust formation, though the structure of SN 1991G could also have been less
favorable for dust formation to occur than in SN 1994W. 

If the steep tail is not primarily powered by radioactive decay we need an 
additional, fast declining, energy source. A light echo seems rather 
improbable as it would produce a spectrum on the tail that is qualitatively
similar to the spectrum at the peak. This is not observed.
A more attractive power source is interaction of the ejecta with 
circumstellar material at $\gtrsim 110$ days.  In \S 4.4 we present
evidence from the narrow line profiles for a thin, dense circumstellar
shell close to the supernova.  The high density we find for the shell
may be sufficient to set up the isothermal-shock-wave scenario
discussed by Grasberg \& Nadyozhin (1986), in which the ejecta's
kinetic energy is transformed to UV and optical emission.  As the
shell is accelerated, the conversion of kinetic energy becomes less
efficient, which could create the steeply-falling light curve.
In this case the photometric point on day 197 gives a firm
upper limit to $\M56$. Assuming no dust, the $\gamma$-rays to be
trapped and  a linear scaling between $R$-luminosity and $\M56$ this 
gives $\M56 \lesssim 0.0026^{+0.0017}_{-0.0011}\ \Msun$. This limit could 
of course be higher if dust is important also in this scenario.

In summary, we suggest that the low luminosity on the light curve tail
is due to a very low mass of ejected nickel.

\subsection{The progenitor and the nature of the collapsed core}

We will now examine whether such a small mass of $^{56}$Ni can be consistent
with the type of progenitor indicated by the light curve up to $\sim 110$ days.
The very low nickel mass we find, if radioactivity is not the prime power source
between days 120 and 197 ($\M56 \lesssim 0.0026^{+0.0017}_{-0.0011}\ \Msun$),
agrees with model calculations only for progenitors in the 
ranges $\MZA \sim 8-10 \Msun$ (\cite{MW88}) and $\MZA \gtrsim 25 \Msun$ 
(\cite{WW95}). If the steep
decline of the light curve tail was due instead to early dust formation, our
more conservative estimate ($\M56 = 0.015^{+0.012}_{-0.008}\ \Msun$) is still
lower than the range $\M56 \sim 0.04-0.20 \Msun$, predicted for progenitors
in the range $\MZA \sim 11-25 \Msun$ (\cite{WW95}), but somewhat higher than
$\M56$ predicted for stars with $\MZA \sim 8-10 \Msun$ (\cite{MW88}). 

Theory thus favors two kinds of progenitor that eject low masses of $^{56}$Ni:
a low mass progenitor, that produces almost no nickel during the explosion, or
a very massive star where the mass-cut (the location in mass 
coordinates between ejected matter and that eventually forming the compact,
central object) is situated sufficiently far out for most of the synthesized 
nickel to be trapped in the compact object. Due to the extensive fall-back of 
material in the latter case, the mass of the compact object may become large 
enough to form a black hole (\cite{WT96}). As a matter of fact, Woosley \& 
Timmes (1996) suggested that the prime observational diagnostic of a
supernova forming a black hole would be a bright supernova that plummeted to 
very low or zero luminosity right after the plateau. 

To estimate the mass of the progenitor we use the early light curve
(i.e., at $t \lesssim 110$~days) of the supernova. At this epoch the
supernova is powered by the release of stored shock energy from the explosion. 
The physics of the plateau phase has been parameterized analytically (Popov 
1993) and fitted to numerical simulations (Litvinova \& Nadyozhin 1985) 
in terms of the explosion energy, $E$, the stellar radius, $R_0$, and 
the mass of the hydrogen envelope at the time of explosion, $M_{\rm env}$. 
Assuming a canonical value for $E$ in the range $(1.0-1.5)\EE{51}$ ergs, 
we use parameterized expressions to estimate $R_0$ and $M_{\rm env}$ for
given $t_{\rm d}$ (the epoch when the plateau ends) and $V$ (the absolute 
visual magnitude before turn-down). 
Taking $t_{\rm d} \sim 110$ days and $V=-17.7\pm0.5$, Popov's expressions
give $M_{\rm env} \sim 7-13 \Msun$ and $R_0 \sim (0.5-2.0)\EE{14}$~cm, while
the numerical models of Litvinova \& Nadyozhin 
give $M_{\rm env} \sim 11-19 \Msun$
and $R_0 \sim (0.4-1.2)\EE{14}$~cm. $M_{\rm env}$ decreases 
with increasing $R_0$ for given $V$ and $E$. 
We thus derive a value of $M_{\rm env}$ which is similar to, or in the case 
of the numerical models, slightly higher than that of SN 1987A 
(\cite{W88}; \cite{SN90}).


Such high values of $M_{\rm env}$ strongly favor the higher mass
range ($\MZA \gtrsim 25 \Msun$) for the progenitor of SN 1994W.
To $M_{\rm env}$ we need to add the mass of the helium- and metal-rich 
core at the time of explosion. For a progenitor with $\MZA = 25 \Msun$,
this is $\sim 9 \Msun$ (\cite{WW95}). To be consistent with our estimate 
of $M_{\rm env}$ for a $\MZA = 25 \Msun$ progenitor, we must therefore allow 
for $3-9 \Msun$ of stellar winds if Popov's expressions are used for the 
plateau, while the numerical models do not require presupernova mass loss. 

A low-mass ($\MZA \sim 8-10 \Msun$) progenitor is possible for SN
1994W, but only if the explosion energy was lower than
the canonical $(1.0-1.5)\EE{51}$ ergs assumed above and the conversion of
explosion energy to radiation was very efficient.
Model B of Falk \& Arnett (1977) shows that a high ($> 30 \%$) efficiency 
is possible if the star is very extended ($R_0 = 9.25\EE{14}$~cm). 
Assuming an explosion energy as low as $3\EE{50}$~ergs, Popov's model 
can give an envelope mass of only $\sim 3.5 \Msun$ for a radius 
of $5\EE{14}$~cm, while the model of Litvinova \& Nadyozhin 
gives $\sim 5.0 \Msun$ for a radius of $3\EE{14}$~cm. 
Both the analytical and numerical estimates are thus consistent with a low
mass ($\MZA \sim 8-10 \Msun$) progenitor, if the progenitor is very extended
and the explosion energy is low. Because a $\MZA \sim 8-10 \Msun$ supernova
only ejects $\M56 \sim 0.002 \Msun$ (\cite{MW88}), 
this type of explosion is consistent only with our lowest $\M56$ (see \S 4.2). 

If SN 1994W resulted from the explosion of an $8 - 10 \Msun$ star,
its compact object should be a neutron star.  If its
progenitor was instead in the higher range, the mass of the
newly-created compact object could become large enough to form a black
hole (\cite{WT96}).  For a $25 \Msun$ progenitor, the mass-cut should 
be $\gtrsim 2.1 \Msun$ (Timmes, Woosley \& Weaver 1996).  
The maximum mass of a neutron star is rather uncertain, depending on the
adopted equation of state. If it is as low as $\sim 1.5-1.6 \Msun$ 
(\cite{BB94}), such a massive progenitor for SN 1994W most likely formed a 
black hole.

\subsection{The narrow lines}
The narrow lines of SN 1994W are not typical for SNe IIP. In \S 3 we
argued that the flux in the lines was too low to affect the light
curve on the plateau, which is why we were able
to apply the light curve analysis in \S 4.3.  Here we discuss the
origin of the narrow lines. As already suggested in \S 3, the lines
are naturally explained in terms of excitation by the hot supernova
continuum.  Another possibility could be excitation caused by
circumstellar interaction of the ejecta with a clumpy wind, as
suggested for SN 1988Z by Chugai \& Danziger (1994).  However, we see
no obvious reason why this excitation mechanism should result in an
excitation of the lines that closely follows the fading of the
supernova continuum as in SN 1994W.

The narrow emission lines were present as early as day $17-18$ 
(\cite{BMB94}, \cite{CLM94}).
H$\alpha$ had a roughly constant P-Cygni profile from then until at
least day 57, the epoch of our last spectrum on the plateau (see
Fig. 3).  The emission part of the P-Cygni profile has a rounded
rather than flat-topped peak, which immediately indicates that the
optical depth in the lines $\gg 1$ (\cite{M78}).  Furthermore, the
steep red side and relatively broad absorption trough observed in 
both $\HI$ and $\FeII$ indicate that the lines were scattering-dominated,
and originated in a rather thin shell ($\Delta R / R < 1$) located
close to the photosphere (see Fransson 1984).  The optical thickness
of the Fe~II lines allows us to put a limit on the density in the shell.  
Assuming a gas temperature of $10^4$~K and Fe~II/H = Fe/H = $4\EE{-5}$
(i.e., the solar value of Fe/H), we use expressions from Mihalas (1978) to
obtain a hydrogen density $n_{\rm H} \gg 3\EE8~v_3~r_{15}^{-1} \cm3$.
Here, $v_3$ is the velocity of the shell in units of $10^3 \kms$
and $r_{15}$ the shell radius in units of $10^{15}$~cm.  We know from the
line profiles that $v \sim 10^3 \kms$, and that the shell radius must
be close to that of the photosphere.  From our blackbody fits (see \S 3) we 
estimate that the photospheric radius on days 31 and 57 
was $\sim 8.5\EE{14}$~cm and $\sim 1.1\EE{15}$~cm, respectively.  This
gives a limit on $n_{\rm H} \gg 3\EE8 \cm3$, quite consistent with the
absence of forbidden lines in the spectrum during the plateau.

The near-constancy of the shell's velocity up to day 57 and its proximity to 
the photosphere allow us to locate the starting radius of the shell, $r_0$, 
assuming it is the same shell that emits the lines at all epochs.  
Using $v_3 = 1$, we obtain $r_0 \sim 6\EE{14}$~cm from both the day 31 and 
day 57 data.  Because the radius of the photosphere is rather uncertain 
($\sim \pm 30\%$), so is $r_0$.

The shell appears to have been quite close to the
supernova around the time of explosion.  We identify two possible
origins.  One possibility is that the shell was created in the
supernova envelope during the explosion.  The formation of such a
shell was suggested by Falk \& Arnett (1977), who actually also
predicted that this should result in narrow P-Cygni profiles.  The
problem with this interpretation is the low velocity of the shell.
Even in their low-explosion-energy, extended
model B, Falk \& Arnett find a shell velocity roughly twice as high as
we find for SN 1994W.

It seems more plausible that the shell was circumstellar, perhaps due to an 
episode of increased mass loss some time $t = (r_0 - R_0) / v_{\rm sh}$ 
prior to the explosion, where $v_{\rm sh}$ is the initial velocity of the 
shell.  This scenario is similar to the model presented by Grasberg \&
Nadyozhin (1986) for SN 1983K.  We find that the radiation from the
supernova can easily accelerate the shell to the observed velocity.
Assuming pure electron scattering, a shell initially at rest is
accelerated up to $v \sim 8.9\EE2~L_{50}~r_{15}^{-2} \kms$, 
where $L_{50}$ is the integrated bolometric luminosity of the supernova 
in units of $10^{50}$~ergs and $r_{15}$ the shell radius in units 
of $10^{15}$~cm.  The velocity can be substantially higher if line 
acceleration is important.  We note that the blue edge of the H$\alpha$ 
absorption shifted slightly to the blue between days 18 and 31, 
corresponding to an increase in shell velocity from $997\pm30 \kms$ 
to $1063\pm15 \kms$ (Fig. 3).  This is consistent with the
supernova radiation continuing to accelerate the shell between these epochs.


In the model of Grasberg \& Nadyozhin (1986), the shell is eventually 
accelerated to the velocity of the supernova ejecta due to circumstellar 
interaction.  However, in reality this may take longer than in 
Grasberg \& Nadyozhin's 1-D models because the slow shell
might break up (\cite{FA77}). In SN 1994W, the disruption of 
the proposed shell by the ejecta could perhaps power the light curve 
between $\sim 120$ and 197 days, as we suggested in \S 4.2.





\section{Conclusions}
We have observed the luminous SN 1994W. We find that the low flux of the
supernova after $\sim 120$ days is evidence of very low mass of $^{56}$Ni
ejected at the explosion, in fact the lowest inferred for any SN II. 

The estimated mass of $^{56}$Ni depends on the interpretation of the unusually 
rapid fading observed in $R$ between days 120 and 197. Either the supernova 
ejected $0.015^{+0.012}_{-0.008}\ \Msun$ of $^{56}$Ni, or the nickel mass 
could have been as low as $\lesssim 0.0026^{+0.0017}_{-0.0011} \Msun$. 
The first possibility requires inefficiency in converting radioactive decay 
energy to $R$ luminosity, and we find that early dust formation is 
the most likely cause for this. Alternatively, the primary power source 
between $120 - 197$ days was not radioactive decay, but perhaps interaction 
with a thin circumstellar shell. 
In this case $\M56 \sim 0.0026^{+0.0017}_{-0.0011} \Msun$ is an upper limit, 
if dust formation was unimportant.

Two types of progenitor seem consistent with the low nickel content and 
high luminosity of the supernova before the end of its plateau phase.
The first possibility is a low mass progenitor with $\MZA \sim 8-10
\Msun$, which had a very extended envelope and a low 
explosion energy, $E \sim 3\EE{50}$~ergs. 
As a low mass progenitor only produces $\sim 0.002 \Msun$ $^{56}$Ni
(\cite{MW88}), consistent only with our lowest value for $\M56$, an
additional power source, presumably circumstellar interaction, is
needed to explain the light curve between $120 - 197$ days for this type
of progenitor. However, the energy budget seems uncomfortably tight for the 
low-mass case as SN 1994W was one of the most luminous supernovae observed. 

A more natural possibility is therefore a high mass progenitor 
with $\MZA \gtrsim 25 \Msun$, which could have lost a few solar masses as 
stellar winds, and exploded with the canonical energy $(1.0-1.5)\EE{51}$ ergs. 
This type of progenitor is consistent with both our estimates of $\M56$.
Such a supernova should have a mass-cut $\gtrsim 2.1 \Msun$, sufficient to 
trap most of the produced $^{56}$Ni, and could have formed a black hole. 

Finally, we find that the narrow lines were formed in a thin, circumstellar
high-density shell close to the supernova, accelerated by the radiation from 
the supernova.


\acknowledgements


We thank P. Rudd, R. Rutten, M. Breare and E. Harlaftis for taking service
observations, E. Zuiderwijk for giving up PATT time to take the 
day 18 spectrum, and M. N\"aslund for taking part in our last observation
of the supernova. We are especially grateful to C. Kozma, B. Leibundgut and 
S. Blinnikov for important discussions, and to L. Szentasko and M. Villi for 
access to unpublished data.  We also thank M. P\'erez-Torres for early 
analysis of the narrow lines and S. Sp\"annare for help with photometry 
measurements. This research was supported by the Swedish Natural Science 
Research Council. JS and PL also acknowledge support from the Swedish National 
Space Board.


\clearpage

%
%
%
\begin{deluxetable}{lrcccl}
\footnotesize
\tablewidth{0pc}
\tablecaption{Supernova 1994W - log of photometric observations}
\tablehead{
\colhead{date}   & \colhead{day}\tablenotemark{1}   & \colhead{telescope/}  &  
\colhead{bandpass}    & \colhead{magnitude}  & \colhead{observer(s)}  \nl	
&
\colhead{}   & \colhead{instrument}   & \colhead{}   &\colhead{}   & 
}

\startdata
\nl
941116.2 & 125 & JKT/TEK4 & V &19.0$\pm$0.2& P. Rudd\nl
 &  &  & R &18.4$\pm$0.5& P. Rudd\nl
941130.3 & 139 & JKT/TEK4 & R & 18.8$\pm$0.5& V. Dhillon\nl
941222.3 & 161 & JKT/TEK4 & R &19.7$\pm$0.6& E. Harlaftis\nl
950120.3 & 190 & JKT/TEK4 & R &20.7$\pm$0.6& E. Harlaftis\nl
950127.3 & 197 & NOT/STANCAM1 & R &21.15$\pm$0.3& J. Sollerman, R. Cumming\nl
960420.0 & 646 & NOT/BROCAM2 & R &$>22$& J. Sollerman, M. N\"aslund \nl

\nl

\tablecomments{Observations at the Jakobus Kapteyn Telescope (JKT)
were taken during service time.}  
\tablenotetext{1}{Days since July 14.0, 1994.}  
\enddata
\end{deluxetable}

%
%
\begin{deluxetable}{lrcccl}
\footnotesize
\tablewidth{0pc}
\tablecaption{Supernova 1994W - log of spectroscopic observations}
\tablehead{
\colhead{date}   & \colhead{day}\tablenotemark{1}   & \colhead{telescope/}  &  
\colhead{wavelength region}   &\colhead{resolution}    & \colhead{observer(s)}  \nl	
&
\colhead{}   & \colhead{instrument}   & \colhead{(\AA)}   &\colhead{(\AA)}   & 
}

\startdata
\nl
940731.9 & 18 & INT/IDS & 6507-6675 &0.25& E. Zuiderwijk \nl
940813.9 & 31 & WHT/ISIS & 4290-8005 &2.7& M. Breare  \nl
940908.9 & 57 & WHT/ISIS & 3300-9360 &11& R. Rutten \nl
940908.9 & 57 & WHT/ISIS & 4255-5060,6345-7150 &2.7& R. Rutten \nl
941112.3 & 121 & INT/IDS & 5485-7370 &6& R. Rutten \nl
950201.5 & 202 & NOT/LDS & 4800-10000 &12& R. Cumming, J. Sollerman  \nl

\nl

\tablecomments{Observations at William Herschel Telescope (WHT) and Isaac
Newton Telescope (INT)  were taken during service time.
The Nordic Optical Telescope (NOT) observations were made by the authors.}
\tablenotetext{1}{Days since July 14.0, 1994.}
\enddata
\end{deluxetable}


\begin{figure}
\plotone{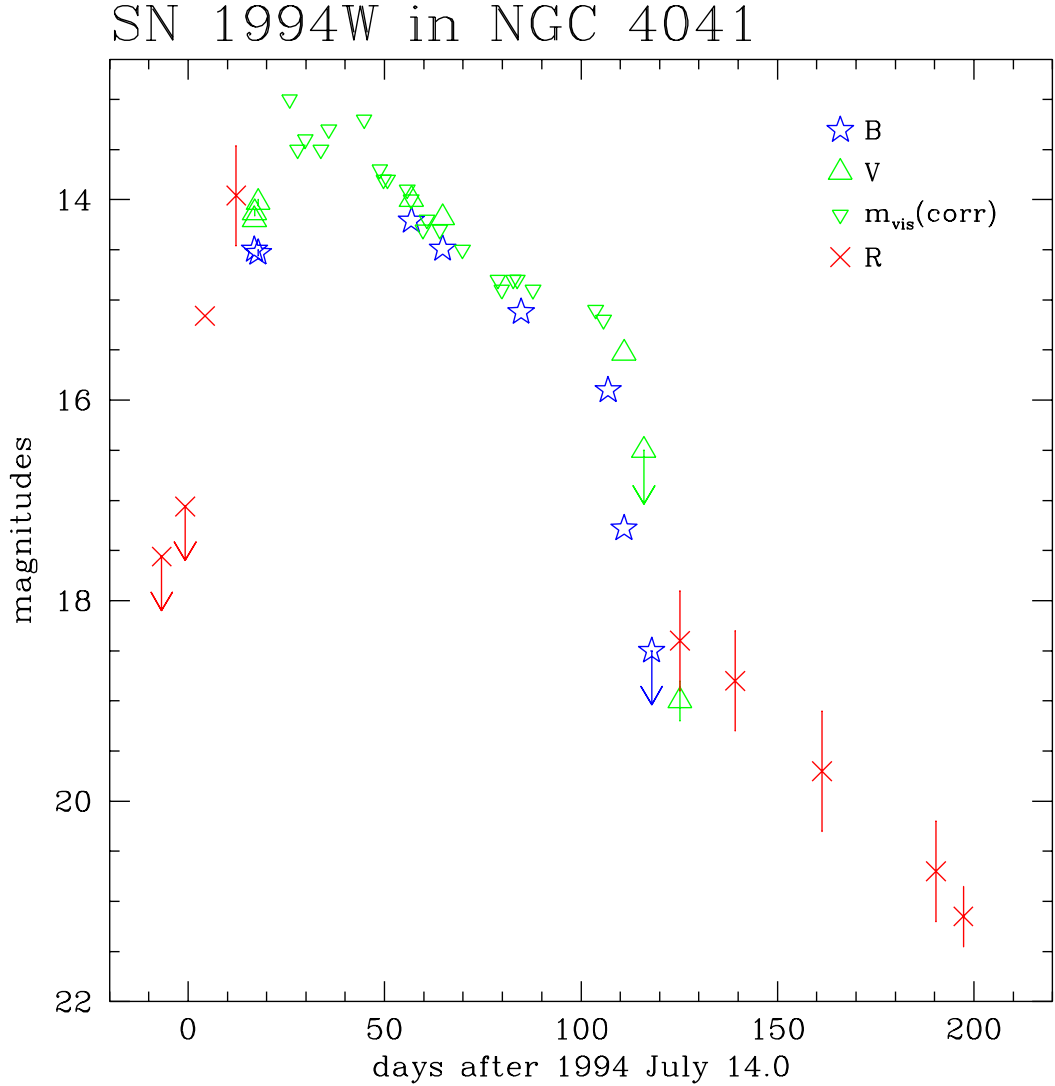}
\caption
{Observed $B$, $V$ and $R$ light curves of SN 1994W.
The data set is a compilation of our own CCD photometry (Table~1),
plus photoelectric and photographic photometry compiled from Tsvetkov
(1995), IAU circulars, and the observations of L.\ Szentasko and
M. Villi (priv. comm.; inverted triangles).  We compared Szentasko's
and Villi's estimates for this supernova, and for SNe 1994D and 1994I
(from the unpublished lists compiled by B. H. Granslo) with
contemporary $V$ and $B-V$ measurements where they were available.
Szentasko's values for all three SNe required a correction of
-0.5$\pm$0.2 mag to match them with $V$.  There was no noticeable
color term.  The corrected magnitudes follow the photoelectric $V$
points well.  The early $R$ points were derived from the relative
photometry of Richmond et al. (1994), using our measurement 
of $R=13.96 \pm 0.15$ for their nearby calibration star (Tsvetkov's 
star~3). 
}
\end{figure}

\begin{figure}
\plotone{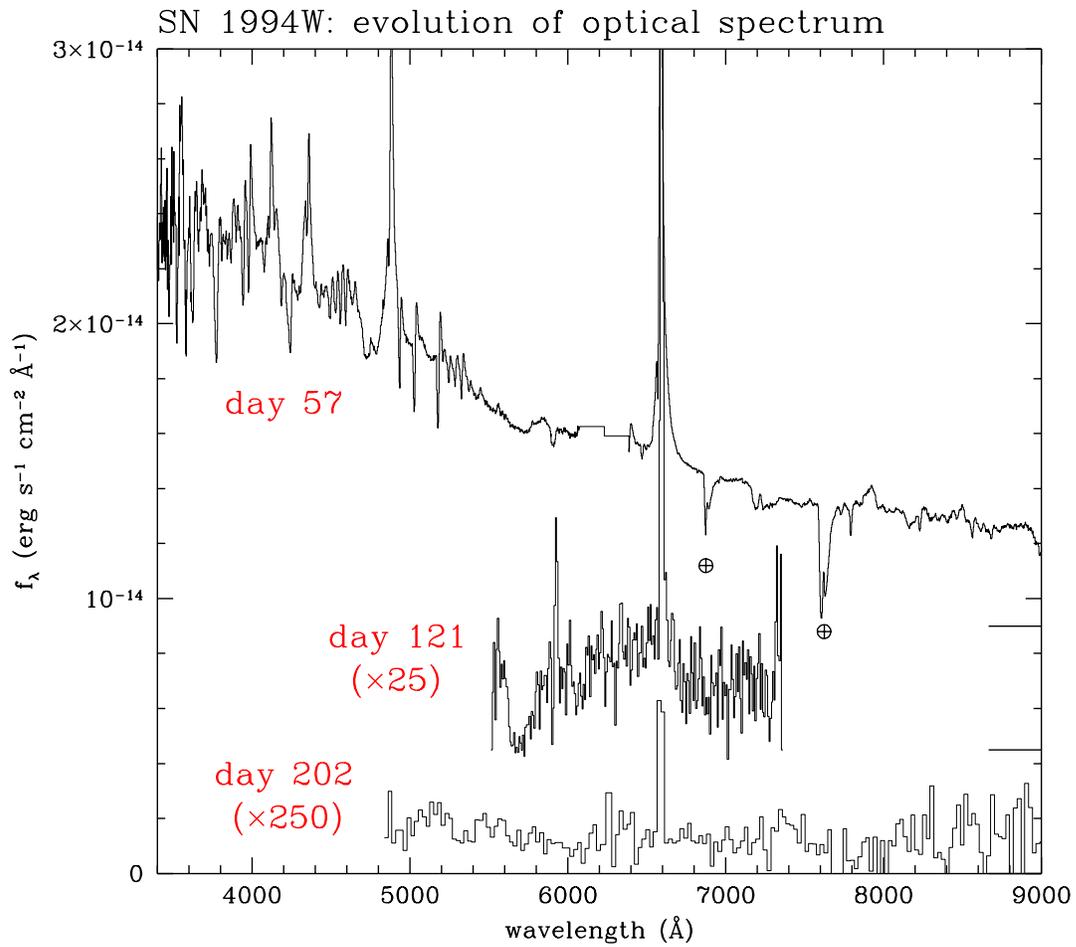}
\caption
{Spectra of SN 1994W at 57, 121 and 202 days after explosion.
The two long tick marks to the right show the level of zero flux at 57 and
121 days.
}
\end{figure}

\begin{figure}
\plotone{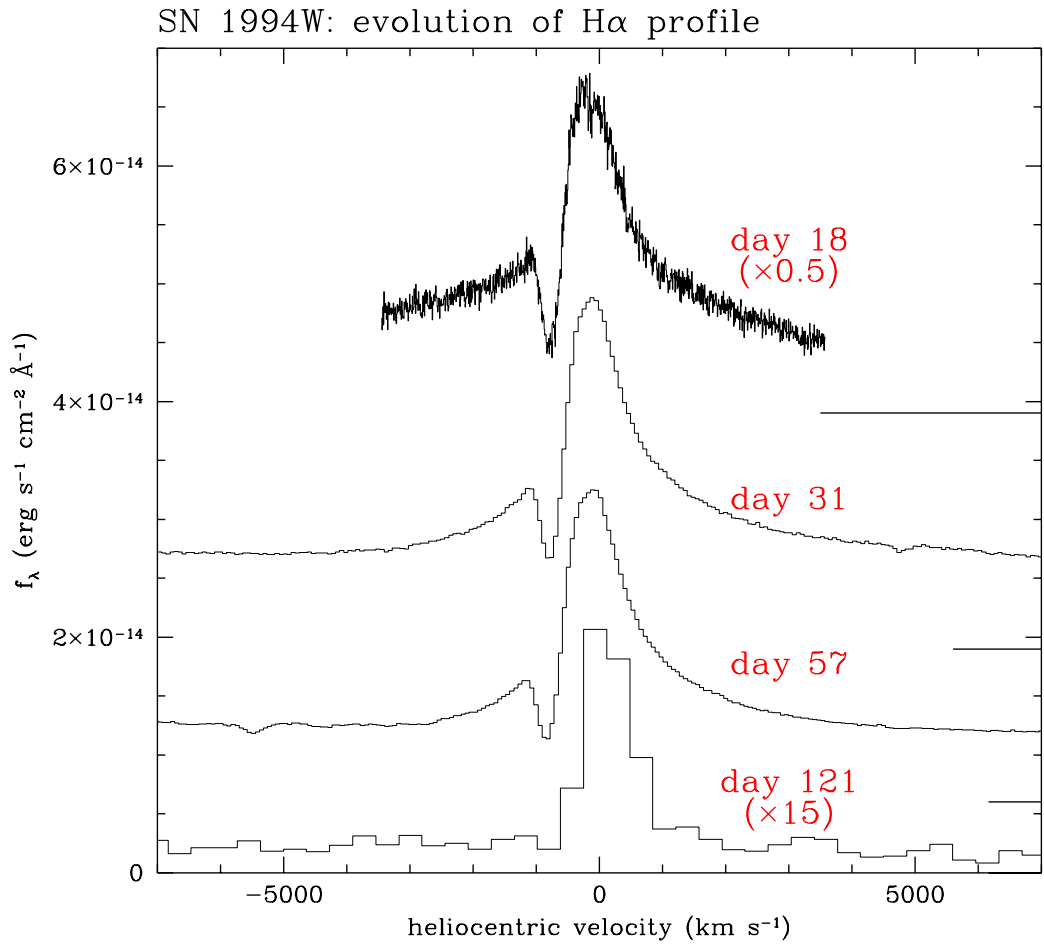}
\caption
{Evolution of the H$\alpha$ line profile of SN 1994W from 
day 18 to day 121. Note the near constancy of the position of the blue edge 
of the absorption. The three long tick marks to the right show the level of
zero flux at 18, 31 and 57 days. 
}
\end{figure}

\begin{figure}
\plotone{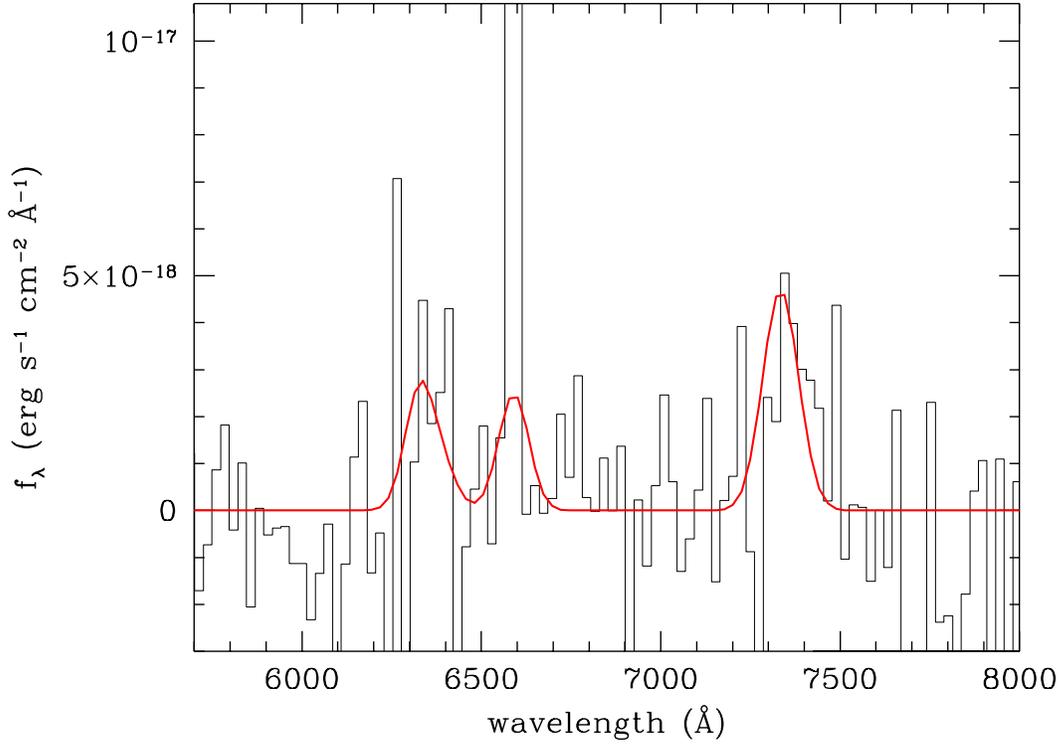}
\caption
{Spectrum of SN 1994W at day 202, with continuum subtracted.  
The solid line (not a fit to the spectrum) shows gaussian profiles of
width 4500 $\kms$ at the positions of 
[O {\sc i}] $\lambda\lambda$6300, 6364, $\Ha$ and 
[Ca {\sc ii}] $\lambda\lambda$7291, 7323 in the ratio 3: 1: 2.5: 3: 2.
}
\end{figure}

\begin{figure}
\plotone{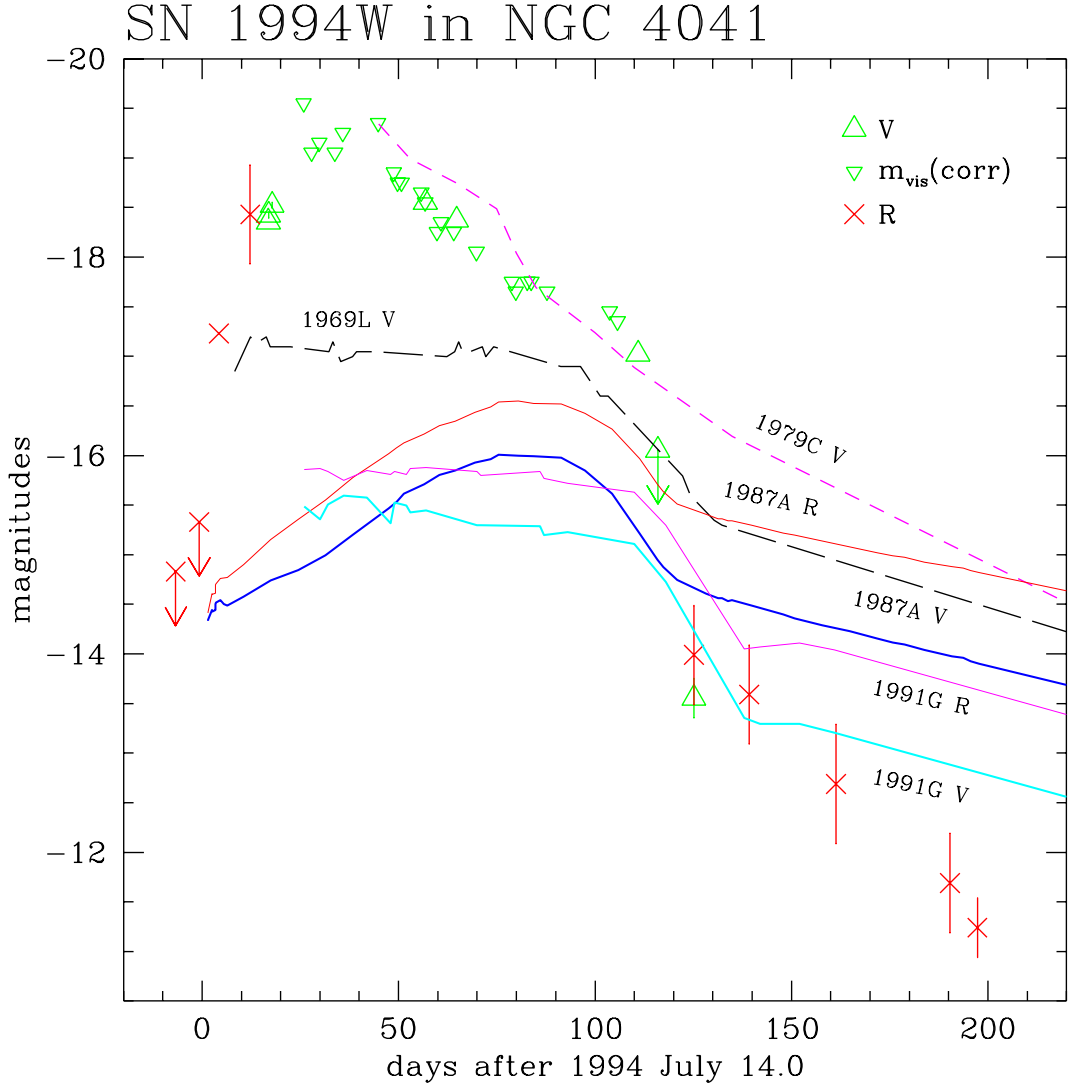}
\caption
{Absolute light curves of SN 1994W compared with selected other
Type II supernovae. We adopt 25.4 Mpc for the distance to SN 1994W 
(taking recession velocity of 1231 $\kms$ from Tully (1988) and 
adopting $H_0=65 \kms$~Mpc$^{-1}$ with the Virgo infall model of 
Kraan-Korteweg [1986]).  Using the relations of Munari \& Zwitter (1997) 
for the strength of interstellar $\NaI$ absorption, we estimated the
extinction towards SN 1994W to be $E(B-V)$=0.17$\pm$0.06.  For the
other SNe the following original data were adopted: SN 1969L (distance
11 Mpc; $A_V$ 0.19 mag; Schmidt et al. 1993); 1979C (16.1 Mpc; 0.45
mag; Ferrarese et al. 1997); 1987A (50 kpc; 0.15 mag; Suntzeff \&
Bouchet 1990); 1990E (21 Mpc; 1.5 mag; 
Schmidt et al. 1993); 1991G (15.5 Mpc; 0.025 mag; Blanton et al. 1995).  
}
\end{figure}

\vfill\eject

\end{document}